\documentclass[superscriptaddress, reprint,nofootinbib, amsmath,amssymb, aps,prb,floatfix,]{revtex4-1}

\usepackage[  ]{graphicx} 
\usepackage{dcolumn} 
\usepackage{bm,subfigure} 

\newcommand{\rmi}{\mathrm{i}} 
\newcommand{\rmm}{\mathrm{m}}

\newcommand{\point}{\raise0.7ex\hbox{.}}

\begin{document}

\preprint{APS/123-QED}

\title{Electrostriction enhancement in metamaterials} 

\author{M. J. A. Smith}
\email{m.smith@physics.usyd.edu.au}
\affiliation{Centre for Ultrahigh bandwidth Devices for Optical Systems (CUDOS), School of Physics, The University of Sydney, NSW 2006, Australia}
\affiliation{ Institute of Photonics and Optical Science (IPOS), School of Physics, The University of Sydney, NSW 2006, Australia}
\author{B. T. Kuhlmey}
\affiliation{Centre for Ultrahigh bandwidth Devices for Optical Systems (CUDOS), School of Physics, The University of Sydney, NSW 2006, Australia}
\affiliation{ Institute of Photonics and Optical Science (IPOS), School of Physics, The University of Sydney, NSW 2006, Australia}
\author{C. Martijn de Sterke}
\affiliation{Centre for Ultrahigh bandwidth Devices for Optical Systems (CUDOS), School of Physics, The University of Sydney, NSW 2006, Australia}
\affiliation{ Institute of Photonics and Optical Science (IPOS), School of Physics, The University of Sydney, NSW 2006, Australia}

\author{C. Wolff}
\author{M. Lapine}
\author{C. G. Poulton}
 \affiliation{Centre for Ultrahigh bandwidth Devices for Optical Systems (CUDOS), School of Mathematical and Physical Sciences, University of Technology Sydney, NSW 2007, Australia }

\date{\today} 

\begin{abstract} \noindent
We demonstrate a controllable enhancement in the  electrostrictive properties of a medium using dilute  composite artificial materials. Analytical expressions  for     the  composite electrostriction   are derived and used to  show  that      enhancement, tunability  and   suppression    can be achieved through a careful choice of   constituent materials.    Numerical examples with  Ag, As$_2$S$_3$, Si  and SiO$_2$  demonstrate that even in a non-resonant regime, artificial materials can bring more than a threefold enhancement in  the electrostriction.

 \begin{description}
\item[PACS numbers]
77.65.Bn, 78.20.H-,81.05.Xj, 42.65.Es
\end{description}

\end{abstract}

\maketitle

\section{Introduction\label{sec:intro}} \noindent
Optoacoustic interactions have gained considerable attention in recent years in the context of nanophotonics \cite{eggleton2013inducing}. One of the strongest and most important of these     is  Stimulated Brillouin Scattering (SBS) \cite{brillouin1922diffusion,mandelstam1926light,chiao1964self}, which is a coherent interaction between the electromagnetic and acoustic fields occurring in an optical waveguide. SBS has been demonstrated in a number of   areas within nanophotonics,   notably in the design of nanoscale devices for Brillouin lasers,    signal processing and   microwave generation \cite{eggleton2013inducing}. The strength of SBS is principally determined by the electrostriction, which is the induced strain arising from an electromagnetic field within the waveguiding material.  The magnitude of the electrostrictive effect, as well as that of the related photo-elastic effect, has widely been considered a property of the material used, and as a consequence, the materials that have been used in SBS studies have been mostly limited to those with naturally large electrostriction constants.

At the same time, it is  well-established in the metamaterials literature that large enhancements in  the  nonlinear properties of a medium can be achieved through the use of composites that have sub-wavelength structural features    \cite{lapine2014colloquium}. Metamaterials have    been used to enhance nonlinear scattering effects such as the Raman effect \cite{agranovich2004linear}, to achieve nonlinear diffraction\cite{segal2015controlling}, and have been  used in  optomechanical systems\cite{aspelmeyer2014cavity} at microwave frequencies \cite{lapine2012magnetoelastic}. However, nonlinear metamaterials have yet to be designed for the enhancement and suppression of electrostriction and photoelasticity, particularly in the optical range. 

\begin{figure}[b]
\centering
\includegraphics[width=0.43\textwidth]{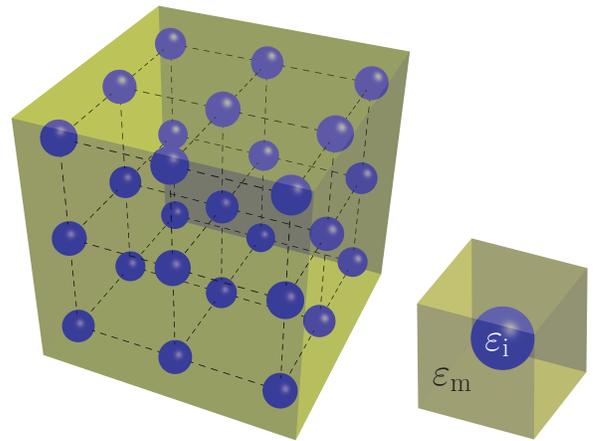} 
\caption{\label{fig:schema} Schematic view of the metamaterial geometry investigated; a primitive cubic array of spheres in a host medium. Inlaid:   fundamental unit cell for a cubic lattice of spheres.}
\end{figure}

In this paper we demonstrate that  artificial materials can be designed for the tuneable enhancement or suppression of  electrostriction. We investigate materials consisting of a dilute suspension of spheres embedded in a dielectric matrix, as presented in Fig. \ref{fig:schema}.  We consider both dielectric and metallic inclusions, and   derive a mixing formula that describes the effective electrostriction of the composite.  The     electrostriction for a selection of practically realisable examples is then evaluated, and used to show that enhancement or  suppression of electrostriction can be achieved.  To our knowledge, we are the first to explore modifications in the   optoacoustic material properties of a medium. It has been  shown previously that even   very simple composite material designs can enhance   the nonlinear susceptibility beyond that of either constituent materials\cite{sipe1992nonlinear}, and therefore, we   expect similar enhancements here with the electrostriction. 

To determine the electrostrictive properties of a composite material, we must first  obtain electrostriction values for all constituent media.  Expressions for these constituents    can differ    depending  on whether   dispersion and loss are incorporated in their derivation, and less obviously,   on other mechanical and thermodynamic assumptions  that are imposed \cite{boyd2003nonlinear,fabelinskii1968molecular,landau1984electrodynamics,rakich2010tailoring}.  These considerations play an important part in determining regimes over which   estimates for the electrostriction are appropriate, and a  discussion of these  relevant approximations can be found in the context of their derivations below.   
 
The outline of this paper is as follows.  In Section \ref{sec:electrogen} we derive a general expression for the  electrostriction,  including the effects of dispersion,    and apply this to uniform dielectrics and metals. In Section \ref{sec:eleccomp}, we obtain the electrostriction for  composite materials.   In Section \ref{sec:numerical} we   consider a series of   practical examples  before concluding remarks   in Section \ref{sec:concl}.

 \section{  Electrostriction   for constituent media}\label{sec:electrogen} \noindent
In this section, we    derive a   general expression for the electrostriction of a homogeneous material. Typically, estimates for the electrostriction of materials are made under the assumption of zero loss and dispersion, zero   shear stress, and that   variations  in the permittivity arise from changes in density alone (i.e. an  isothermal process)  \cite{boyd2003nonlinear,landau1984electrodynamics,stratton2007electromagnetic}.      In a generalisation of the standard procedure, we   incorporate the effects of  dispersion in our derivation. We  begin by considering  the electromagnetic energy density   \cite{landau1984electrodynamics}
\begin{equation}
\label{eq:endendisp}
u = \frac{1}{2} \varepsilon_0 \frac{\partial (\omega \varepsilon_\mathrm{r})}{\partial \omega} |\mathbf{E}|^2,
\end{equation}
where $\varepsilon_0$ denotes the free-space permittivity, $\varepsilon_\mathrm{r}$ is the relative permittivity of the material, $\omega$ is the frequency and $\mathbf{E}  $ is  the   electric field. The change in energy density corresponding to a small change in the density $\rho$ is therefore 
 \begin{subequations}
\begin{equation}
\Delta u = \frac{1}{2} \varepsilon_0 \frac{\partial }{\partial \rho} \left[ \frac{\partial (\omega \varepsilon_\mathrm{r})}{\partial \omega} \right] |\mathbf{E}|^2 \Delta \rho,
\end{equation}
 where $\Delta$   denotes an infinitesimal quantity. Assuming an isothermal process, this change in the internal energy  can be equated to the work done $W$ per unit volume $V$ by    
\begin{equation}
\label{eq:thermo}
\Delta W = P \frac{\Delta V}{V} = -P \frac{\Delta \rho}{\rho},
\end{equation}
 \end{subequations}
where  $P$ is the induced hydrostatic pressure, to obtain 
\begin{equation}
\label{eq:pressg}
 P =    -\frac{1}{2}\varepsilon_0\gamma  |\mathbf{E}|^2 ,
\end{equation}
and we define the electrostriction  parameter
\begin{equation}
\label{eq:gammametal}
\gamma = \rho \frac{\partial^2 ( \varepsilon_\mathrm{r} \omega) }{\partial \rho \partial \omega},
\end{equation}
as a nondimensional measure of the induced electrostrictive stress. From a microscopic perspective, the pressure field in  \eqref{eq:pressg}  can be understood   as arising from     ionic movements in the material lattice induced by a Lorentz force \cite{nelson1971theory,mueller1935theory}. From   \eqref{eq:gammametal}, we obtain expressions for the  electrostriction of both dielectric and metallic media which are used in our composite model shown later.

 \subsection{Electrostriction for dielectric media}\label{sec:diel} \noindent
For a dielectric medium that is nondispersive and lossless, the electrostriction parameter   \eqref{eq:gammametal} simplifies to the familiar form \cite{boyd2003nonlinear}
\begin{subequations}
 \begin{equation}
\label{eq:GAMMA}
  \gamma  = \rho \frac{\partial \varepsilon_\mathrm{r}}{\partial \rho}.
 \end{equation}
It is then usual to express this     in terms of other well-known material response tensors for practical evaluation. For example, for isotropic and homogeneous materials, \eqref{eq:GAMMA} is given by  
\begin{equation}
\label{eq:gammaps}
\gamma =   \frac{1}{3} \varepsilon_\mathrm{r}^2 \left( p_{11} + 2 p_{12} \right),  
\end{equation}
\end{subequations}
 where $p_{ij}$ denotes the elastooptic coefficients of the medium \cite{rakich2010tailoring,melloni1998direct,landau1984electrodynamics}.     These $p_{ij}$ coefficients are well-tabulated for a range of materials,  and a selection of values for dielectric solids are presented in Table \ref{tab:table1} for reference.  However, to our knowledge  no experimental data has been published on elastooptic coefficients   for metallic media\cite{stakhin1998electrostriction}, and so we now consider an  estimate  for the $\gamma$  of metals.  

\begin{table}[b]
\caption{\label{tab:table1}%
Material parameters for a selection of dielectric materials at specified wavelengths,  $\gamma$ from Eq. \eqref{eq:gammaps}.}
\begin{ruledtabular}
\begin{tabular}{lcccccc}
Material &$\lambda$ (nm)& $\varepsilon_\mathrm{r}$ & $p_{11}$ & $p_{12}$ & $\gamma$& Ref\\
\hline
 SiO$_2$ &$663$& $2.12$ & 0.12 & 0.27& 1.00 &\cite{weber2002handbook} \\
As$_2$S$_3$  &$1150$& $6.06$ & 0.31 & 0.30& 11.1 & \cite{weber2002handbook} \\
Si  &$3390$& $11.8$ & -0.09 & 0.02& -2.77 & \cite{biegelsen1974photoelastic}
\end{tabular}
\end{ruledtabular}
\end{table}

\subsection{Electrostriction for metallic media}\label{sec:elecmet} \noindent 
For metals, we return to   \eqref{eq:gammametal}, and use  a simplified   Drude model  for the permittivity\cite{bohren2008absorption}
\begin{equation}
\label{eq:drude}
\varepsilon_\mathrm{r}   = 1 - \frac{\omega_\mathrm{p}^2}{    \omega^2},
\end{equation}
 to obtain a form for $\gamma$ which is  useful for practical evaluation. Here,   
$\omega_\mathrm{p}^2 =  {q^2 N}/{(\varepsilon_0 m_\mathrm{e})}$ is the square of the plasma frequency,  $q$ is the electric charge, $m_\mathrm{e}$ is the effective mass of a constituent electron,     $N= \rho / m$ is the number density, and $m$ is the mass density of the metal. 
Subsequently  \eqref{eq:gammametal} and \eqref{eq:drude} give    the estimate  
\begin{equation}
\label{eq:gammaDm}
\gamma^\mathrm{DM} =  \frac{   \omega_p^2     }{\omega^2}.  
\end{equation}
We note that in the derivation of the metallic $\gamma$ above, we have   neglected dissipation effects, which is consistent with the isothermal   assumption made in the derivation of \eqref{eq:gammametal}. The validity of this assumption is considered in the results section, with a discussion of attenuation. 

Having derived  evaluable expressions  for dielectrics and metals, we   now   proceed to the electrostriction of    composite materials.

\section{Electrostriction for composite materials}\label{sec:eleccomp} \noindent
 In this section, we derive   $\gamma$ for our composite material using the Maxwell-Garnett (MG)  model. The effective permittivity given by this  model is valid   for a dilute   array of spheres embedded in a host material, and has the form     \cite{milton2002theory} 
\begin{equation}
\label{eq:MG}
\varepsilon_\mathrm{r}^\mathrm{ } =  \varepsilon_ {\mathrm{m}}  + \frac{3 \varepsilon_ {\mathrm{m}}   ( \varepsilon_ {\mathrm{i}} - \varepsilon_ {\mathrm{m}})f}{( \varepsilon_ {\mathrm{i}} + 2 \varepsilon_ {\mathrm{m}}) - ( \varepsilon_ {\mathrm{i}} - \varepsilon_ {\mathrm{m}})f }.
\end{equation}
Here  $\varepsilon_\mathrm{i,m}$   denotes the relative permittivities of the  constituent     materials, and we define the filling fraction
\begin{equation}
\label{eq:fillfrac}
f = \frac{V_ {\mathrm{i}}}{V_ {\mathrm{i}} + V_ {\mathrm{m}}},
\end{equation}
where $V_\mathrm{i,m}$ represent corresponding   volumes. The subscript $\mathrm{i}$   denotes the inclusion and $\mathrm{m}$   denotes the matrix (constrained by the boundaries of the unit cell), as shown by the fundamental cell   in Fig. \ref{fig:schema}.  The only additional condition for   \eqref{eq:MG} in our analysis is that the periodic array must be suitably subwavelength. An established  rule of thumb is that the period of the lattice must be at least 10 times smaller than the optical wavelength in the material. To demonstrate this    in a practical context,  a wavelength in the material of 1 $\mu  \mathrm{m}$ and a filling fraction of  $f = 15\%$ would correspond to  an array period  of $100 \, \mathrm{nm}$ and a spherical radius of  $33 \, \mathrm{nm}$.

 To begin, we consider a fully nondispersive model for the composite electrostriction.

\subsection{Nondispersive model} \noindent
Under the assumption that all constituent materials are nondispersive, the $\gamma$ expression \eqref{eq:gammametal} reduces to the form given in   \eqref{eq:GAMMA}. Consequently, from the MG model  we write
\begin{equation}
\gamma^\mathrm{ND} = \rho\left[       \frac{\partial \varepsilon_\mathrm{r}^\mathrm{ }}{\partial \varepsilon_ {\mathrm{m}} } \frac{\partial \varepsilon_ {\mathrm{m}} }{\partial \rho} +\frac{\partial \varepsilon_\mathrm{r}^\mathrm{ }}{\partial \varepsilon_ {\mathrm{i}}  } \frac{\partial \varepsilon_ {\mathrm{i}} }{\partial \rho} + \frac{\partial \varepsilon_\mathrm{r}^\mathrm{ }}{\partial f}  \frac{\partial f}{\partial \rho} \right],
\label{eq:esccr}
\end{equation}
where from \eqref{eq:MG}  we have the partial derivatives    
\begin{subequations}
\label{eq:firstderivs}
\begin{align}
  \frac{\partial \varepsilon_\mathrm{r}^\mathrm{ }}{\partial \varepsilon_ {\mathrm{m}} } &= 
  \frac{\left[ (\varepsilon_\mathrm{i} + 2\varepsilon_\mathrm{m} )^2 + 2f (\varepsilon_\mathrm{i} - \varepsilon_\mathrm{m})^2 \right] (1-f)}{\left[  ( \varepsilon_ {\mathrm{i}} + 2 \varepsilon_ {\mathrm{m}}) - ( \varepsilon_ {\mathrm{i}}  - \varepsilon_ {\mathrm{m}}) f \right]^2},\\
    \frac{\partial \varepsilon_\mathrm{r}^\mathrm{ }}{\partial \varepsilon_ {\mathrm{i}} } &= \frac{ \left[ (  \varepsilon_ {\mathrm{i}} + 2 \varepsilon_ {\mathrm{m}}  ) - ( \varepsilon_ {\mathrm{i}} - \varepsilon_ {\mathrm{m}}  )\right]    ^2 f}{\left[  ( \varepsilon_ {\mathrm{i}} + 2 \varepsilon_ {\mathrm{m}}) - ( \varepsilon_ {\mathrm{i}}  - \varepsilon_ {\mathrm{m}}) f \right]^2},\\
      \frac{\partial \varepsilon_\mathrm{r}^\mathrm{ }}{\partial f }   &= \frac{3 \varepsilon_ {\mathrm{m}} ( \varepsilon_ {\mathrm{i}} - \varepsilon_ {\mathrm{m}})( \varepsilon_ {\mathrm{i}} + 2 \varepsilon_ {\mathrm{m}}) }{\left[  ( \varepsilon_ {\mathrm{i}} + 2 \varepsilon_ {\mathrm{m}}) -  ( \varepsilon_ {\mathrm{i}}  - \varepsilon_ {\mathrm{m}})f \right]^2}.
\end{align}
\end{subequations}
However, to evaluate the remaining three derivatives in  \eqref{eq:esccr}, we need to consider the mechanical response of the metamaterial to the   induced pressure field \eqref{eq:pressg}.

Since shear stress is   omitted  in our model, the matrix and inclusions do not undergo deformations, but instead  experience a compression in order to preserve hydrostatic equilibrium. This is seen mathematically by    stating that  perturbations to the material pressure fields   remain continuous     across the boundary of the sphere
\begin{equation}
\Delta P_ {\mathrm{i}}\big|_{\partial\Omega} = \Delta P_ {\mathrm{m}}\big|_{\partial\Omega},
\label{eq:elbc}
\end{equation}
where $P_\mathrm{i,m}$ denotes     the   pressure fields  and $\partial \Omega$ is the boundary of the inclusion. 
We can then evaluate    Taylor series for the  constituent volumes  $V_\mathrm{i,m}$ with respect to $P_\mathrm{i,m}$ to obtain 
\begin{subequations}
\begin{align}
\Delta V_ {\mathrm{i}}&= \frac{\partial V_ {\mathrm{i}}}{\partial P_ {\mathrm{i}}} \Delta P_ {\mathrm{i}},  \\
\Delta V_ {\mathrm{m}} &= \frac{\partial V_ {\mathrm{m}}}{\partial P_ {\mathrm{m}}} \Delta P_ {\mathrm{m}},
\end{align}
\end{subequations}
 and express \eqref{eq:elbc} in the form
\begin{align}
\label{eq:BCdensity}
\frac{\Delta V_\mathrm{i} }{V_ {\mathrm{i}}\beta_ {\mathrm{i}}}\bigg|_{\partial \Omega} &= \frac{\Delta  V_\mathrm{m}  }{V_ {\mathrm{m}} \beta_ {\mathrm{m}}}\bigg|_{\partial \Omega} ,
\end{align}
where we have introduced the compressibility  constant
\begin{equation}
\beta = -\frac{1}{V} \frac{\partial V}{\partial P} = K^{-1},
\end{equation}
  and $K$ denotes the  bulk modulus. 
Integrating both sides of \eqref{eq:BCdensity} we obtain the   interface condition
\begin{equation}
V_ {\mathrm{m}}= A \left[  V_ {\mathrm{i}} \, \right]^{\beta_ {\mathrm{m}}/\beta_ {\mathrm{i}}}, 
\label{eq:vevi}
\end{equation}
 for some constant $A$, which describes the compressive response of our composite. With this condition, and using  the definition for the composite density
\begin{align}
\rho &=  \rho_ {\mathrm{i}}f + \rho_ {\mathrm{m}} (1 - f) = \frac{m_ {\mathrm{i}} + m_ {\mathrm{m}}}{V_ {\mathrm{i}} + V_ {\mathrm{m}}}, 
\end{align}
we evaluate the remaining three derivatives to obtain
\begin{subequations}
\label{eq:3eqns2}
\begin{align}
\label{eq:crossd1}
 \frac{\partial \varepsilon_ {\mathrm{m}}}{\partial \rho} &=    \frac{\partial \varepsilon_ {\mathrm{m}}}{\partial \rho_ {\mathrm{m}}} \frac{\partial \rho_ {\mathrm{m}}}{\partial V_ {\mathrm{i}}} \frac{\partial V_ {\mathrm{i}}}{\partial \rho} = \frac{\gamma_ {\mathrm{m}}  }{\rho  }  \frac{\beta_ {\mathrm{m}}   }{\beta_ {\mathrm{c}}  } ,\\
\label{eq:crossd2}
 \frac{\partial \varepsilon_ {\mathrm{i}}}{\partial \rho} &=  \frac{\partial \varepsilon_ {\mathrm{i}}}{\partial \rho_ {\mathrm{i}}} \frac{\partial \rho_ {\mathrm{i}}}{\partial V_ {\mathrm{i}}} \frac{\partial V_ {\mathrm{i}}}{\partial \rho}= \frac{\gamma_ {\mathrm{i}} }{\rho } \frac{\beta_ {\mathrm{i}}   }{\beta_ {\mathrm{c}}  }, \\
 \label{eq:dfdp}
 \frac{\partial f}{\partial \rho} &= \frac{\partial f}{\partial V_ {\mathrm{i}}} \frac{\partial V_ {\mathrm{i}}}{\partial \rho} =  \frac{ f (1 - f)  }{\rho  }  \frac{(\beta_ {\mathrm{m}} - \beta_ {\mathrm{i}})}{ \beta_ {\mathrm{c}}},
\end{align}
\end{subequations}
where analogously to  \eqref{eq:GAMMA} we introduce $\gamma_ {\mathrm{m}} = \rho_ {\mathrm{m}} \partial \varepsilon_ {\mathrm{m}} \slash \partial \rho_ {\mathrm{m}}$ and $\gamma_ {\mathrm{i}} = \rho_ {\mathrm{i}}\partial \varepsilon_ {\mathrm{i}} \slash \partial \rho_ {\mathrm{i}}$ as the electrostriction values of the constituent media, and $\beta_ {\mathrm{c}} = \beta_ {\mathrm{i}} f + \beta_ {\mathrm{m}} (1-f)$ denotes the volume-averaged compressibility over the unit cell.  Consequently, the nondispersive electrostriction for our composite    is   given by
\begin{widetext}
\begin{multline}
  \label{eq:GammaMG}
\gamma^\mathrm{ND} =   \frac{\beta_ {\mathrm{i}}f}{\beta_ {\mathrm{c}}}\left[ \frac{ ( \varepsilon_ {\mathrm{i}} + 2 \varepsilon_ {\mathrm{m}} )  - ( \varepsilon_ {\mathrm{i}} - \varepsilon_ {\mathrm{m}}) }{ ( \varepsilon_ {\mathrm{i}} + 2 \varepsilon_ {\mathrm{m}}) -  ( \varepsilon_ {\mathrm{i}} - \varepsilon_ {\mathrm{m}}) f }\right]^2  \gamma_ {\mathrm{i}} + 
 \frac{\beta_ {\mathrm{m}}(1-f)}{\beta_ {\mathrm{c}}} \left[ \frac{( \varepsilon_ {\mathrm{i}} + 2 \varepsilon_ {\mathrm{m}} )^2 + 2f ( \varepsilon_ {\mathrm{i}} - \varepsilon_ {\mathrm{m}} )^2 }{ \left[ ( \varepsilon_ {\mathrm{i}} + 2 \varepsilon_ {\mathrm{m}}) - ( \varepsilon_ {\mathrm{i}} - \varepsilon_ {\mathrm{m}}) f\right]^2}\right] \gamma_ {\mathrm{m}} \\ 
+ \underbrace{\frac{(\beta_ {\mathrm{m}} - \beta_ {\mathrm{i}})f(1-f)}{\beta_ {\mathrm{c}}} \left[ \frac{3 \varepsilon_ {\mathrm{m}} ( \varepsilon_ {\mathrm{i}} - \varepsilon_ {\mathrm{m}} ) ( \varepsilon_ {\mathrm{i}} + 2 \varepsilon_ {\mathrm{m}} ) }{\left[ ( \varepsilon_ {\mathrm{i}} + 2 \varepsilon_ {\mathrm{m}}) - ( \varepsilon_ {\mathrm{i}} - \varepsilon_ {\mathrm{m}}) f\right]^2} \right]}_{\mbox{\footnotesize artificial electrostriction}}, 
 \end{multline}
\end{widetext}
which is a weighted linear function of the constituent electrostriction values $\gamma_\mathrm{i}$ and $\gamma_\mathrm{m}$ plus a new artificial electrostriction term (highlighted).
The latter term can be understood by considering the limit   $\gamma_\mathrm{i} = \gamma_\mathrm{m} =0$; if the two materials have different compressibility values,  then compression leads to a change in the filling fraction $f$, which, if   $\varepsilon_\mathrm{i} \neq \varepsilon_\mathrm{m}$, alters the effective dielectric constant \eqref{eq:MG}.  Another interesting feature of \eqref{eq:GammaMG} is the   second-order pole present in all terms at
\begin{equation}
\label{eq:fres}
f = ( \varepsilon_ {\mathrm{i}} + 2 \varepsilon_ {\mathrm{m}})/( \varepsilon_ {\mathrm{i}} - \varepsilon_ {\mathrm{m}}),
 \end{equation}
  giving  a theoretically infinite value for the composite electrostriction. However, this resonance  can only be obtained with a   change in sign for either $\varepsilon_\mathrm{i}$ or $\varepsilon_\mathrm{m}$ for dilute, positive $f$.  A discussion of the asymptotic behaviour of \eqref{eq:GammaMG} with respect to $\beta_\mathrm{i,m}$ and $\varepsilon_\mathrm{i,m}$ is presented in Appendix \ref{app:1} for completeness. 
  
Next we consider the composite $\gamma$ expression  when dispersion is included.

\subsection{Dispersive corrections} \noindent 
In this section, we     incorporate dispersion in the derivation of  the composite $\gamma$. We   begin by returning to    \eqref{eq:gammametal}, which from the MG model \eqref{eq:MG},  has the form
\begin{multline}
\label{eq:gammacompdispgen}
\gamma = \rho\left[       \frac{\partial \varepsilon_\mathrm{r}^\mathrm{ }}{\partial \varepsilon_ {\mathrm{m}} } \frac{\partial \varepsilon_ {\mathrm{m}} }{\partial \rho} +\frac{\partial \varepsilon_\mathrm{r}^\mathrm{ }}{\partial \varepsilon_ {\mathrm{i}}  } \frac{\partial \varepsilon_ {\mathrm{i}} }{\partial \rho} + \frac{\partial \varepsilon_\mathrm{r}^\mathrm{ }}{\partial f}  \frac{\partial f}{\partial \rho} \right] \\
+ \rho \omega \frac{\partial}{\partial \rho} \left\{       \frac{\partial \varepsilon_\mathrm{r}^\mathrm{ }}{\partial \varepsilon_ {\mathrm{m}} } \frac{\partial \varepsilon_ {\mathrm{m}} }{\partial \omega} +\frac{\partial \varepsilon_\mathrm{r}^\mathrm{ }}{\partial \varepsilon_ {\mathrm{i}}  } \frac{\partial \varepsilon_ {\mathrm{i}} }{\partial \omega} + \frac{\partial \varepsilon_\mathrm{r}^\mathrm{ }}{\partial f}  \frac{\partial f}{\partial \omega} \right\}.
\end{multline}
 This composite expression is then  decomposed in the form $\gamma = \gamma^\mathrm{ND} + \gamma^\mathrm{D}$, where   $\gamma^\mathrm{ND}$ and $\gamma^\mathrm{D}$ represent  the nondispersive and  dispersive contributions, respectively. The  nondispersive contribution is given by the first three terms of        \eqref{eq:gammacompdispgen} and has been evaluated in the previous section as \eqref{eq:GammaMG}, where we introduce  the substitution $\gamma_{\mathrm{i},\mathrm{m}} = \gamma_{\mathrm{i},\mathrm{m}}^\mathrm{ND}$   therein. 

Next, we evaluate the remaining terms in \eqref{eq:gammacompdispgen}, and note that  we have   
\begin{equation}
\frac{\partial f }{ \partial \omega} =0,
\end{equation}
   as  all mechanical parameters, such as $\beta$ and $\rho$, are independent of the optical frequency. Accordingly, we decompose the dispersive term $\gamma^\mathrm{D}$ into a  matrix and inclusion   contribution $ \gamma^\mathrm{D} =   \gamma^\mathrm{D}_{\mathcal{M}} +  \gamma^\mathrm{D}_{\mathcal{I}}$  and obtain
\begin{multline}
\label{eq:GMD}
\gamma^\mathrm{D}_{\mathcal{M}}   =\rho \omega \frac{\partial \varepsilon_ {\mathrm{m}} }{\partial \omega}   \frac{3}{\left[ ( \varepsilon_ {\mathrm{i}} + 2 \varepsilon_ {\mathrm{m}}) -  ( \varepsilon_ {\mathrm{i}} - \varepsilon_ {\mathrm{m}}) f\right]^3}  \cdot \\
\left[ \frac{\partial \varepsilon_\mathrm{i}}{\partial \rho} \left\{6 \varepsilon_\mathrm{i}\varepsilon_\mathrm{m} f(1-f) \right\}   -  \frac{\partial \varepsilon_\mathrm{m}}{\partial \rho} \left\{6 \varepsilon_\mathrm{i}^2 f(1-f) \right\}  \right. \\ \left. + \frac{\partial f}{\partial \rho} \left\{   \varepsilon_\mathrm{i}^3 (1-f) - 3 \varepsilon_\mathrm{i}^2 \varepsilon_\mathrm{m} f - 2 \varepsilon_\mathrm{m}^3 (f+2) - 6 \varepsilon_\mathrm{i}\varepsilon_\mathrm{m}^2(1-f)\right\}       \right] \\
+\frac{\beta_ {\mathrm{m}}(1-f)}{\beta_ {\mathrm{c}}} \left[ \frac{( \varepsilon_ {\mathrm{i}} + 2 \varepsilon_ {\mathrm{m}} )^2 + 2f ( \varepsilon_ {\mathrm{i}} - \varepsilon_ {\mathrm{m}} )^2 }{ \left[ ( \varepsilon_ {\mathrm{i}} + 2 \varepsilon_ {\mathrm{m}}) - f( \varepsilon_ {\mathrm{i}} - \varepsilon_ {\mathrm{m}}) \right]^2}\right] \gamma_ {\mathrm{m}}^D,
\end{multline}
and
\begin{multline}
\label{eq:GID}
\gamma^\mathrm{D}_{\mathcal{I}}  =   \rho \omega \frac{\partial \varepsilon_ {\mathrm{i}} }{\partial \omega}   \frac{9  \varepsilon_\mathrm{m} }{\left[ ( \varepsilon_ {\mathrm{i}} + 2 \varepsilon_ {\mathrm{m}}) -  ( \varepsilon_ {\mathrm{i}} - \varepsilon_ {\mathrm{m}}) f\right]^3}  \cdot \\
\left[ \frac{\partial \varepsilon_\mathrm{m}}{\partial \rho} \left\{ 2 \varepsilon_\mathrm{i}    f(1-f)  \right\} - \frac{\partial \varepsilon_\mathrm{i}}{\partial \rho}  \left\{ 2  \varepsilon_\mathrm{m}  f(1-f)   \right\} \right. \\ 
\left. + \frac{\partial f}{\partial \rho} \left\{    \varepsilon_\mathrm{m}  \left[ ( \varepsilon_\mathrm{i}  +2  \varepsilon_\mathrm{m}  ) + f( \varepsilon_\mathrm{i}   -  \varepsilon_\mathrm{m})  \right]   \right\} \right] \\
 + \frac{\beta_ {\mathrm{i}}f}{\beta_ {\mathrm{c}}}\left[ \frac{ ( \varepsilon_ {\mathrm{i}} + 2 \varepsilon_ {\mathrm{m}} )  - ( \varepsilon_ {\mathrm{i}} - \varepsilon_ {\mathrm{m}}) }{ ( \varepsilon_ {\mathrm{i}} + 2 \varepsilon_ {\mathrm{m}}) - f( \varepsilon_ {\mathrm{i}} - \varepsilon_ {\mathrm{m}}) }\right]^2  \gamma_ {\mathrm{i}}^D. 
\end{multline}
The remaining derivatives  with respect to $\rho$ in \eqref{eq:GMD} and \eqref{eq:GID} are given in \eqref{eq:3eqns2} where we use the substitutions $\gamma_{\mathrm{i},\mathrm{m}} = \gamma_{\mathrm{i},\mathrm{m}}^\mathrm{ND}$   therein.

In summary, the  expression for the dispersive composite electrostriction is given by
\begin{equation}
\label{eq:gammawdispcomp}
\gamma = \gamma^\mathrm{ND}+ \gamma^\mathrm{D}_{\mathcal{M}}   +\gamma^\mathrm{D}_{\mathcal{I}},  
\end{equation}
where $\gamma_{\mathrm{i},\mathrm{m}} = \gamma_{\mathrm{i},\mathrm{m}}^\mathrm{ND}$  has been substituted appropriately.  Using the full  definitions \eqref{eq:gammametal} for the constituents  $\gamma_\mathrm{i,m}$, \eqref{eq:gammawdispcomp} has an identical structure to that presented for the nondispersive expression in \eqref{eq:GammaMG}, except now the artificial electrostriction term is modified by additional terms. These dispersive    contributions \eqref{eq:GMD} and \eqref{eq:GID} feature the same MG resonance \eqref{eq:fres} as before, but with a contribution from a third-order pole. This suggests that the omission of dispersion can, in certain instances,  have considerable influence on the result for composite $\gamma$.
  
  \section{Numerical examples}\label{sec:numerical} \noindent
In this section, we     investigate         the   composite electrostriction expressions \eqref{eq:GammaMG} and \eqref{eq:gammawdispcomp} for our structure, using combinations of  different materials. We accompany this investigation  with an analysis of the losses for these designs, which is necessary for realistic applications. Accordingly,  we return to the MG model   \eqref{eq:MG}  and define the attenuation  length 
\begin{equation}
\label{eq:attenC}
\alpha_\mathrm{L} = \left[\frac{4\pi}{\lambda} \mathrm{Im} \left( \sqrt{\varepsilon_\mathrm{r}} \right) \right]^{-1},
\end{equation}
which we emphasise, is completely independent from the electrostriction analysis.   From \eqref{eq:attenC},   a threshold of $\alpha_\mathrm{L}  \geqslant 0.1 \mathrm{mm}$ is imposed as a tolerance for omitting dissipation effects, which is also a typical interaction length for  SBS.
 
\begin{figure}[t]
\centering
\subfigure[ \label{fig:gss}]{
\includegraphics[scale=0.33]{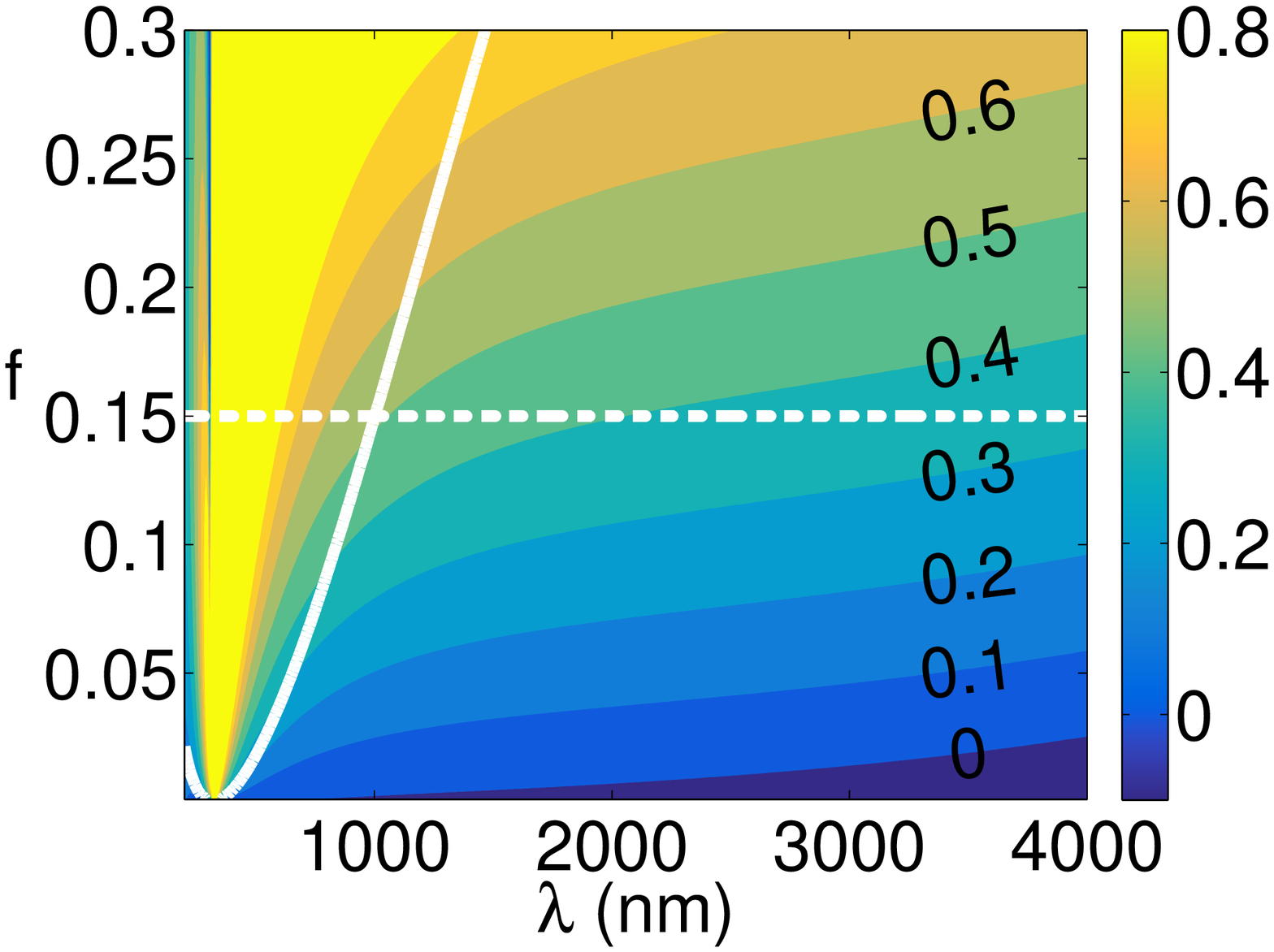}}
\subfigure[  \label{fig:ess} ]{
\includegraphics[scale=0.33]{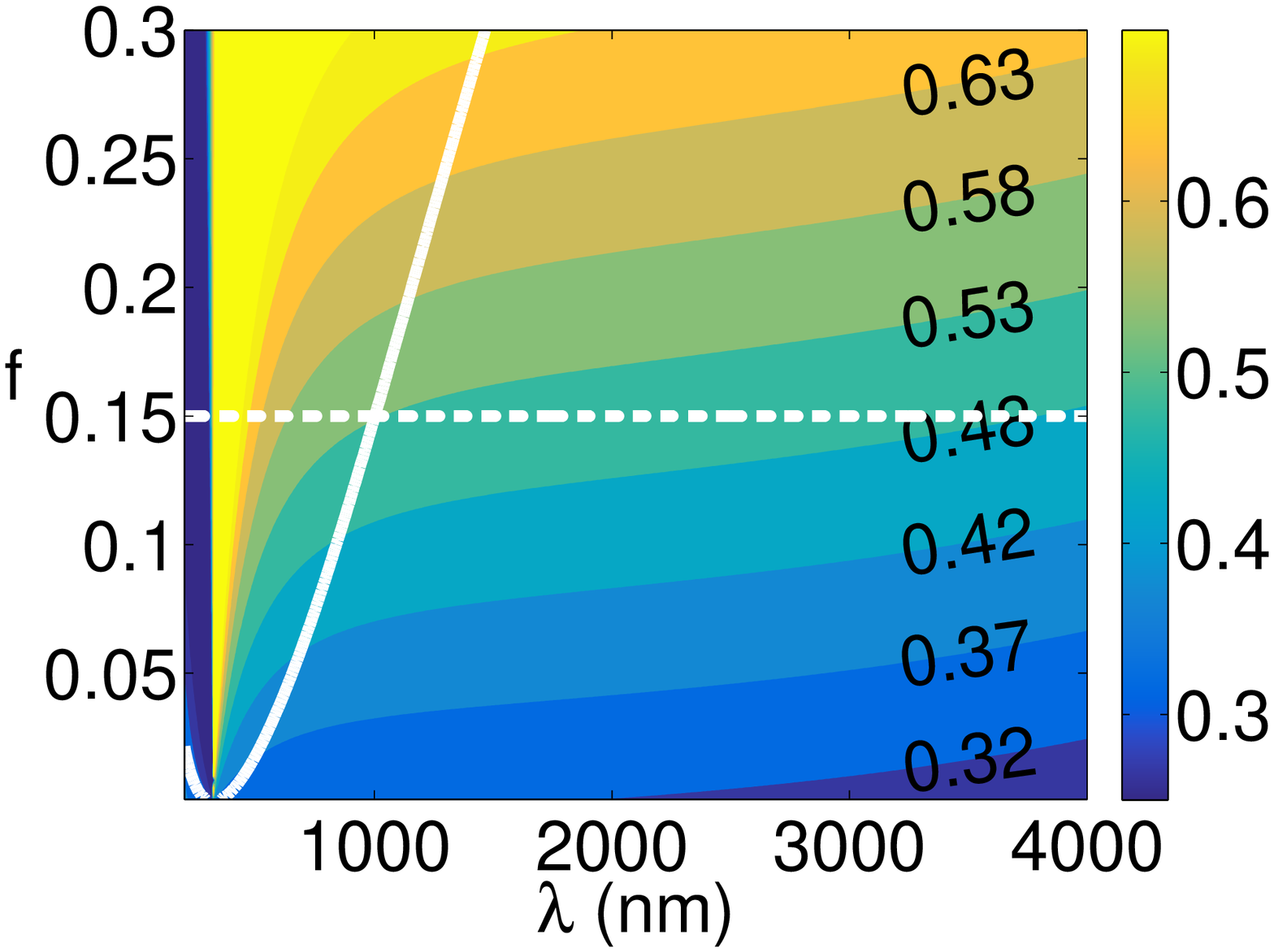}}
\caption{\label{fig:gss_ess}Contour plots of  \subref{fig:gss}    $\log_{10}|\mathrm{Re} (\gamma)|$  from  \eqref{eq:gammawdispcomp}, and    \subref{fig:ess} $\log_{10}|\mathrm{Re} (\varepsilon_\mathrm{r})|$ from   \eqref{eq:MG}, for an array of Ag spheres embedded in a  SiO$_2$ matrix, against filling fraction $f$ and wavelength $\lambda$. A diluteness threshold of $f=15\%$ (dashed white curve) and an attenuation length threshold of $\alpha_\mathrm{L} = 0.1 \, \mathrm{mm}$ (solid white curve) are also shown.}
\end{figure}

 We begin by investigating the composite electrostriction \eqref{eq:gammawdispcomp}  for  a cubic array of silver\cite{ordal1985optical} spheres   embedded in a  silica\cite{malitson1965interspecimen} matrix, where we use \eqref{eq:gammaDm} for $\gamma_\mathrm{i}$. In Fig. \ref{fig:gss} we present a contour plot of    $\log_{10}|\mathrm{Re} (\gamma)|$ over the wavelength range $ 350 \, \mathrm{nm} \leqslant \lambda \leqslant 4000 \,  \mathrm{nm} $ for   filling fraction  $0 \leqslant f \leqslant 0.3$.

A striking feature of this figure is the region corresponding to $\log_{10}|\mathrm{Re}(\gamma)|>0.8$, which contains the permittivity resonance \eqref{eq:fres}. This region simply denotes   $\gamma$ values over a cut-off threshold, which is introduced  to ensure that features of the contour plot are not dominated by the singularity in \eqref{eq:gammawdispcomp}. We note that the  extremely strong enhancements in $\gamma$ courtesy of  \eqref{eq:fres} are     associated   with   strong attenuation \eqref{eq:attenC}, and we highlight this by superposing a solid    white curve over these contours, which represents an attenuation length threshold of $\alpha_\mathrm{L} = 0.1 \, \mathrm{mm}$ (where to the right of this curve we have  longer $\alpha_\mathrm{L} $, and to the left, a region of shorter lengths). 

Also shown is a dashed white curve, which represents our diluteness threshold of $f = 15 \%$. Accordingly, inside the region bound by these two curves (the region of validity (ROV)), we find a maximum composite electrostriction value of $\gamma = 3.27 $ at $(\lambda,f) = (1003 \, \mathrm{nm},0.15) $, which corresponds to the intersection of the $\alpha_\mathrm{L}$ and $f$ curves. This point gives an enhancement factor of   $3.36$  relative to the electrostriction    for the silica background at the same wavelength. It is also clear from these contours that the electrostriction is tuneable over a wide wavelength interval, but that these enhancements  are ultimately constrained by the diluteness requirement of the MG  model. 
  
In Fig. \ref{fig:ess} we present  a contour plot  of the  effective  permittivity over the same ($\lambda,f)$ range, where the plasmon resonance is   clearly visible. For our maximum electrostriction value  at $(\lambda,f) = (1003 \, \mathrm{nm}, 0.15)$, we   have a composite permittivity of $\varepsilon_\mathrm{r} =  3.4 +   0.003 \rmi$, which is an enhancement factor of $1.6$ relative to the background value of  $ \varepsilon_ {\mathrm{m}} = 2.10 $ at the same wavelength. As one would expect, this  contour plot      features similar curvature to that of $\gamma$, and a low degree of frequency dependence within the ROV ($1.9 < \mathrm{Re}(\varepsilon_\mathrm{r}) < 3.4$).  
 
\begin{figure}[t]
\centering
\subfigure[ \label{fig:gcs}]{
\includegraphics[scale=0.33]{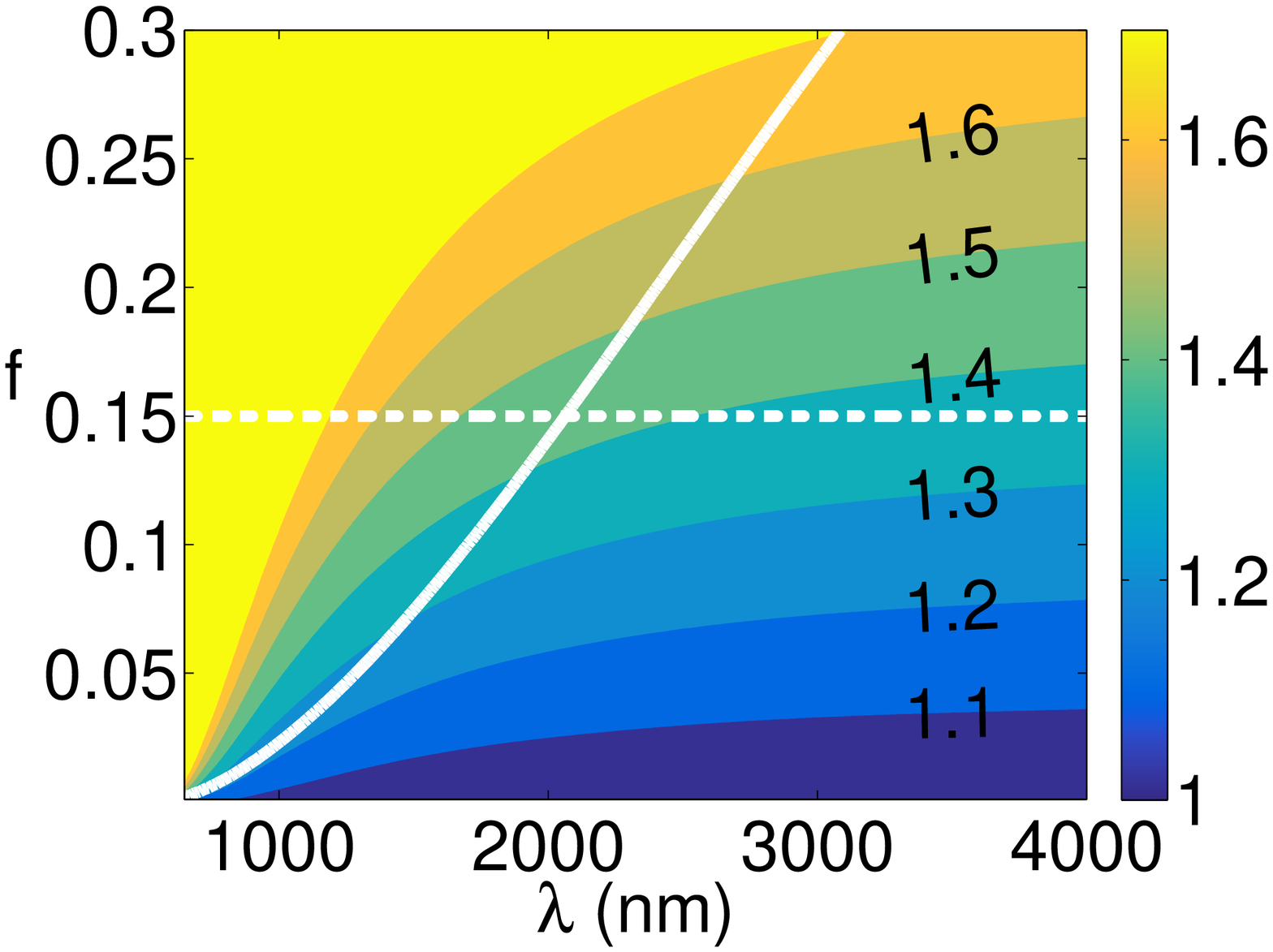}}
\subfigure[  \label{fig:ecs} ]{
\includegraphics[scale=0.33]{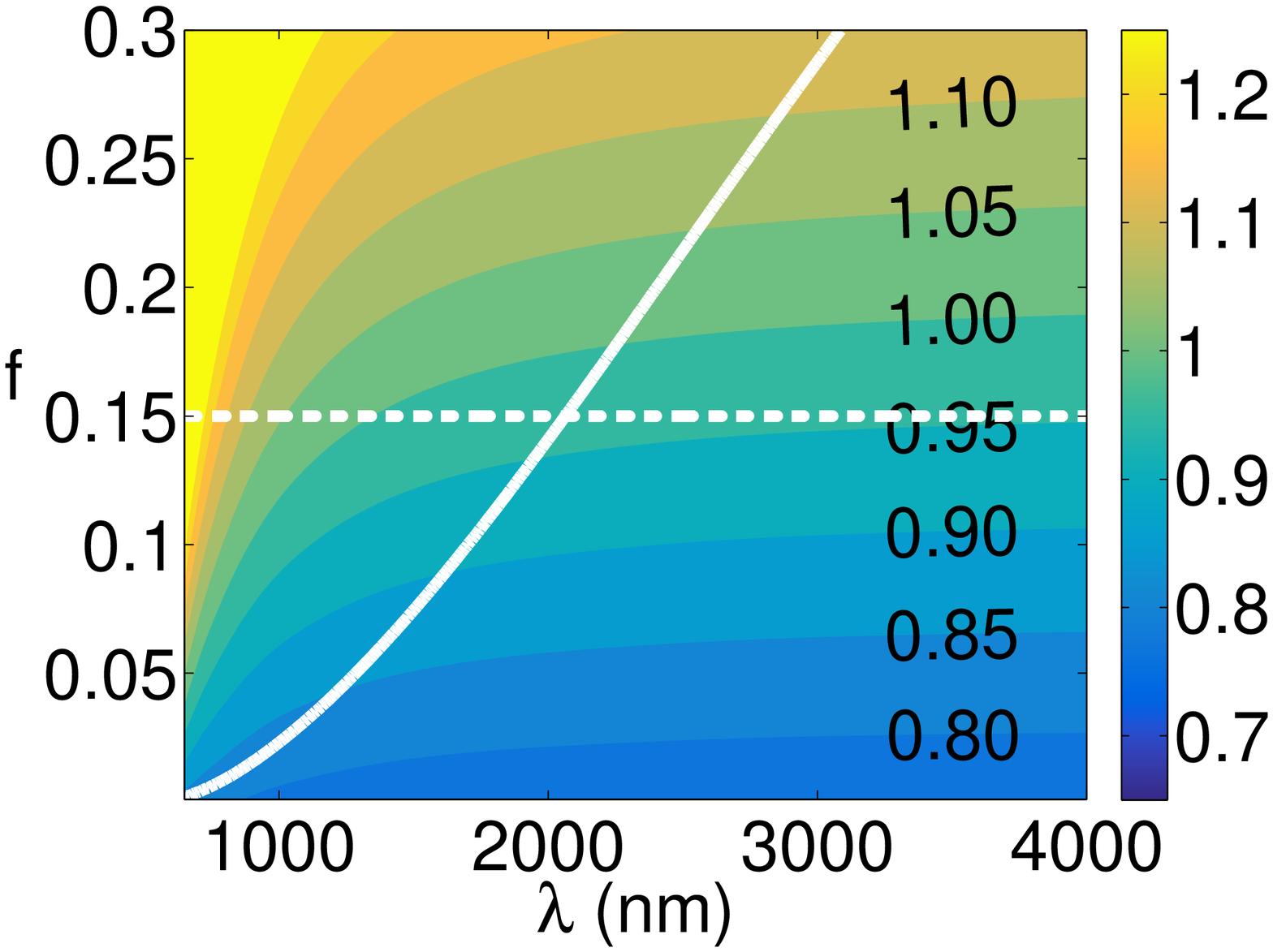}}
\caption{\label{fig:gcs_ecs}Contour plots of  \subref{fig:gss}    $\log_{10}|\mathrm{Re} (\gamma)|$  from  \eqref{eq:gammawdispcomp}, and    \subref{fig:ess} $\log_{10}|\mathrm{Re} (\varepsilon_\mathrm{r})|$ from   \eqref{eq:MG}, for an array of Ag spheres embedded in a  As$_2$S$_3$ matrix, against filling fraction $f$ and wavelength $\lambda$. A diluteness threshold of $f=15\%$ (dashed white curve) and an attenuation length threshold of $\alpha_\mathrm{L} = 0.1 \, \mathrm{mm}$ (solid white curve) are also shown.}
\end{figure}

We now consider silver spheres embedded in a chalcogenide\cite{amorphmat} matrix (amorphous As$_2$S$_3$). In Fig. \ref{fig:gcs} we present the composite $\gamma$ for this configuration,  and observe   qualitatively similar behaviour to   the previous example for a silica matrix  in Fig. \ref{fig:gcs}. The primary difference here is the much more restrictive $\alpha_\mathrm{L}$ threshold, which now extends to much longer wavelengths.

If one searches inside the   ROV constrained by our $\alpha_\mathrm{L}$ and $f$ bounds, we discover a maximum electrostriction value of $\gamma = 27.4$ at $(\lambda,f) = (2064 \, \mathrm{nm},  0.15)$, corresponding to the intersection of the $\alpha_\mathrm{L}$ and $f$ curves as before, with an enhancement factor of   2.63 (c.f., $\gamma_ {\mathrm{m}} = 10.44$).  Fig \ref{fig:ecs} reveals that this coordinate point has an effective permittivity value of $\varepsilon_\mathrm{r} = 9.3 + 0.01\rmi$. This corresponds to a similar permittivity enhancement  factor as the previous example (c.f., $ \varepsilon_ {\mathrm{m}} = 5.89$). A slightly higher level of frequency dependence is observed   in the ROV also ($5.8 < \mathrm{Re}(\varepsilon_\mathrm{r}) < 9.3$).

For these examples, we find that      the composite electrostriction expression \eqref{eq:gammawdispcomp} gives a $10-20\%$ increase in the maximum electrostriction  value compared to   the  nondispersive expression \eqref{eq:GammaMG}. This suggests that the omission of dispersion can give rise to a small but non-negligible correction to the composite electrostriction. Furthermore, a similar investigation  with Au spheres embedded in these   matrix materials reveals a comparable level of enhancement to Ag.

\begin{figure}[t]
\centering
\subfigure[ \label{fig:gsis}]{
\includegraphics[scale=0.33]{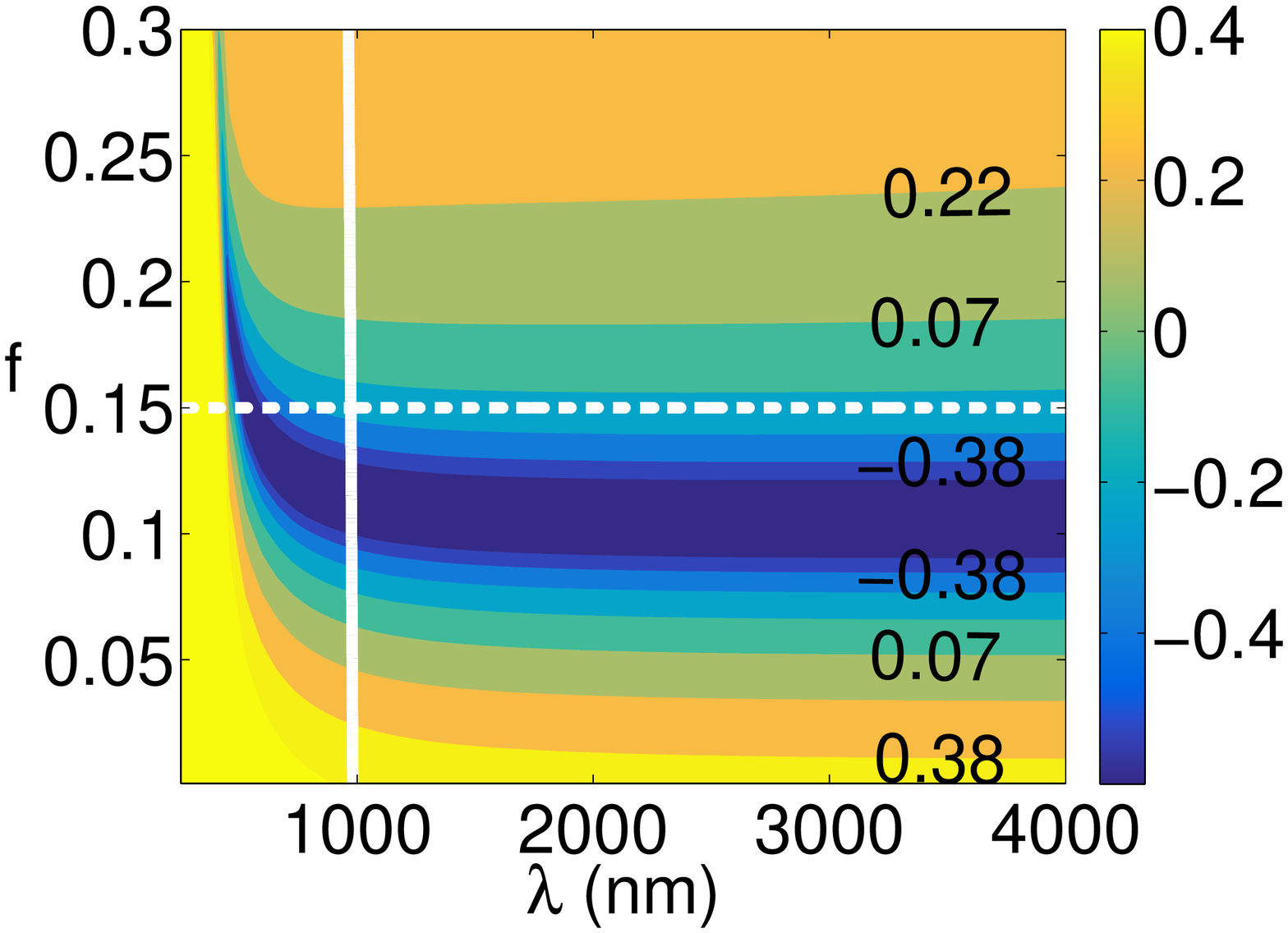}}
\subfigure[  \label{fig:esis} ]{
\includegraphics[scale=0.33]{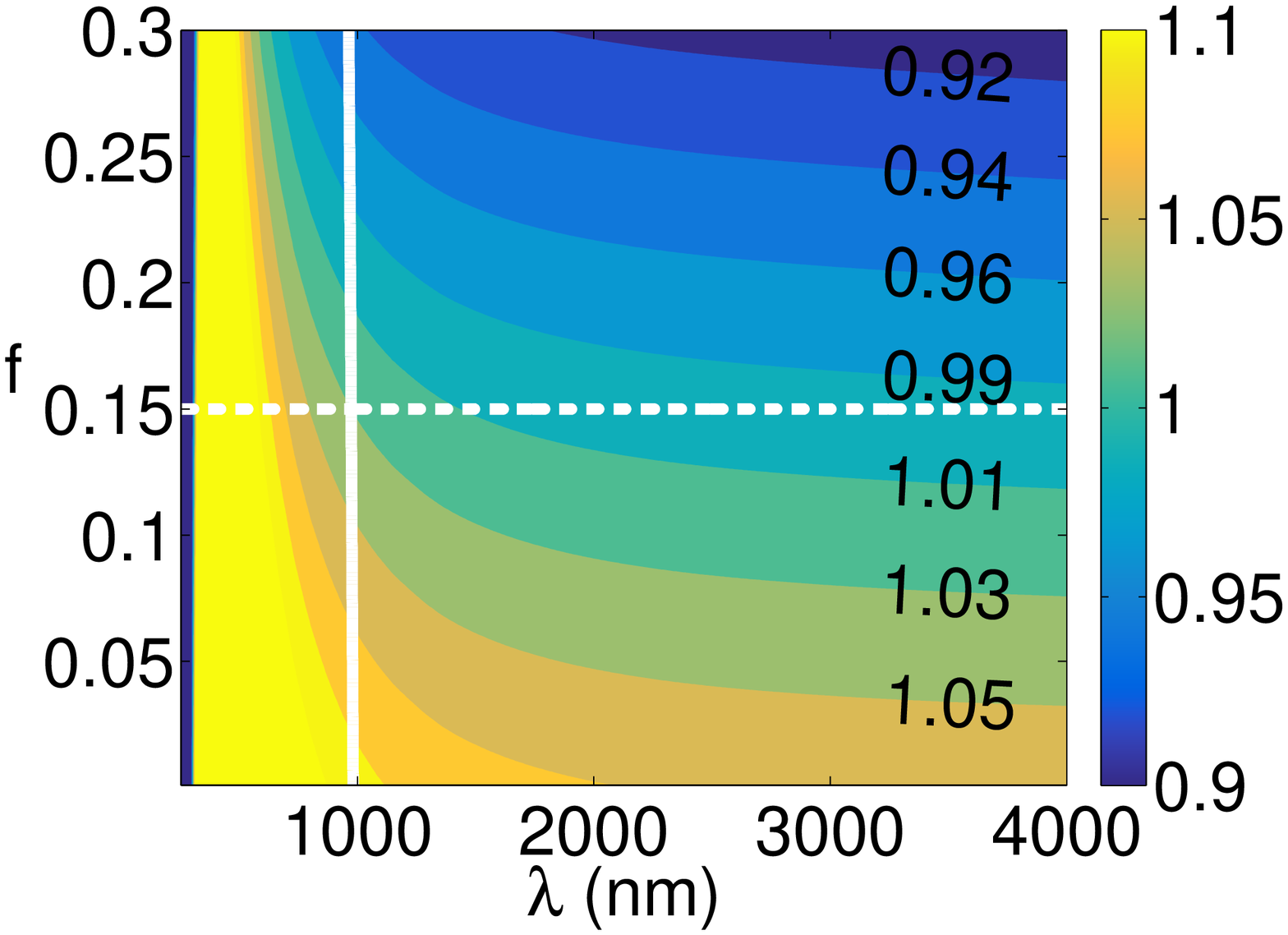}}
\caption{\label{fig:gsis_esis}Contour plots of  \subref{fig:gss}    $\log_{10}|\mathrm{Re} (\gamma)|$  from  \eqref{eq:GammaMG}, and    \subref{fig:ess} $\log_{10}|\mathrm{Re} (\varepsilon_\mathrm{r})|$ from   \eqref{eq:MG}, for an array of SiO$_2$ spheres embedded in a  Si matrix, against filling fraction $f$ and wavelength $\lambda$. A diluteness threshold of $f=15\%$ (dashed white curve) and an attenuation length threshold of $\alpha_\mathrm{L} = 0.1 \, \mathrm{mm}$ (solid white curve) are also shown. }
\end{figure}

In  Fig. \ref{fig:gsis} we consider  $\log_{10}|\mathrm{Re} (\gamma)|$ from \eqref{eq:GammaMG} for silica spheres embedded in a silicon\cite{green2008self} matrix. This figure exhibits strong  frequency dependence for $\lambda < 1000 \, \mathrm{nm}$ (courtesy of  a material resonance for Si at $\lambda \approx 370 \, \mathrm{nm}$) and a near-horizontal arc of   zero electrostriction which spans   the entire ROV.  That is, this metamaterial design can completely suppress electrostriction over an exceptionally wide frequency range. For this particular composite  the attenuation length threshold is reached at approximately $\lambda = 1000 \, \mathrm{nm}$. In Fig. \ref{fig:esis} we present  a contour plot of $\log_{10}|\mathrm{Re} (\epsilon_r)|$ for completeness, which exhibits reassuringly minimal frequency dependence over the ROV.

\begin{figure}[t]
\centering
\subfigure[ \label{fig:fig7}]{
\includegraphics[scale=0.33]{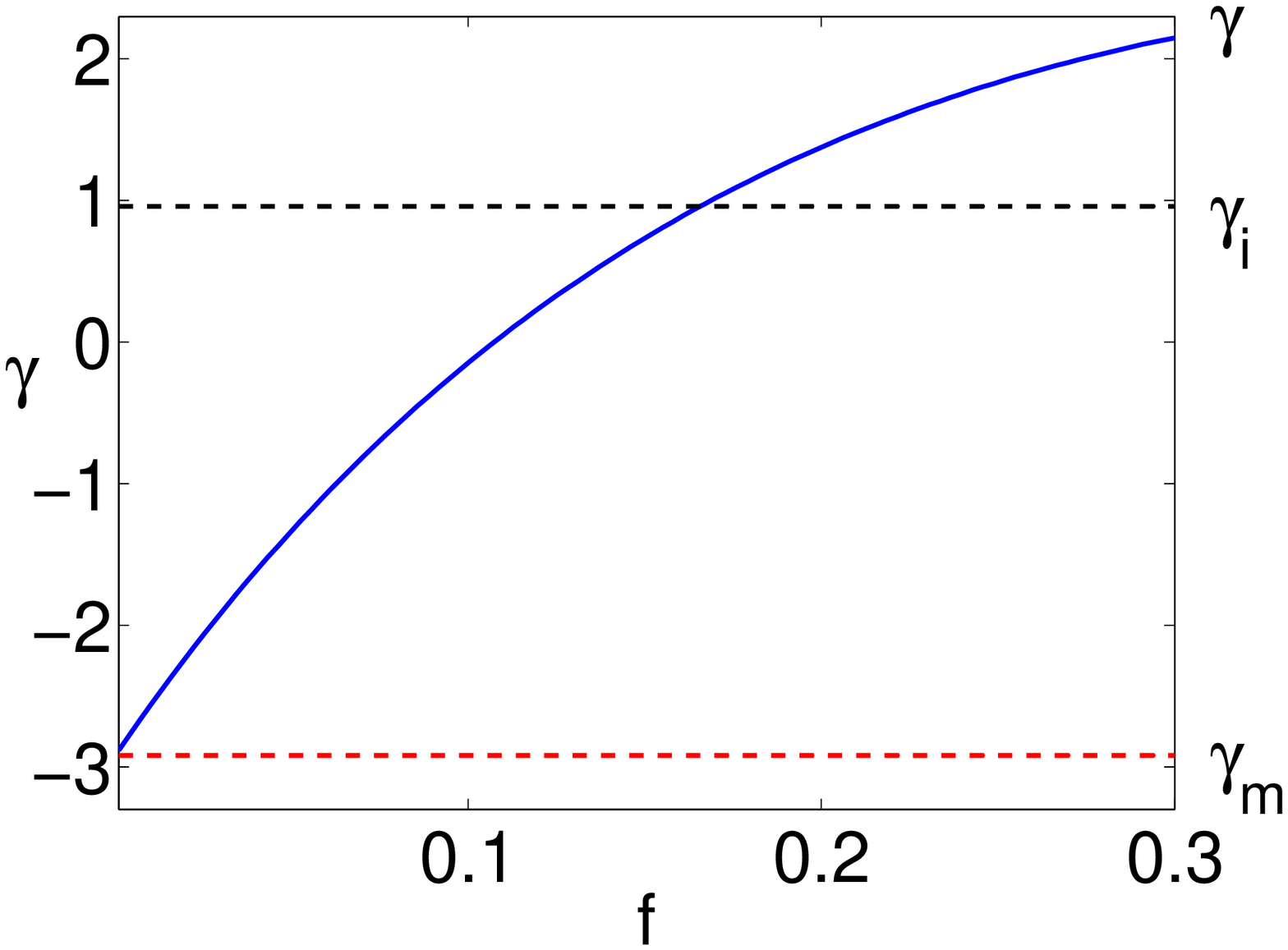}}
\subfigure[  \label{fig:fig7b} ]{
\includegraphics[scale=0.33]{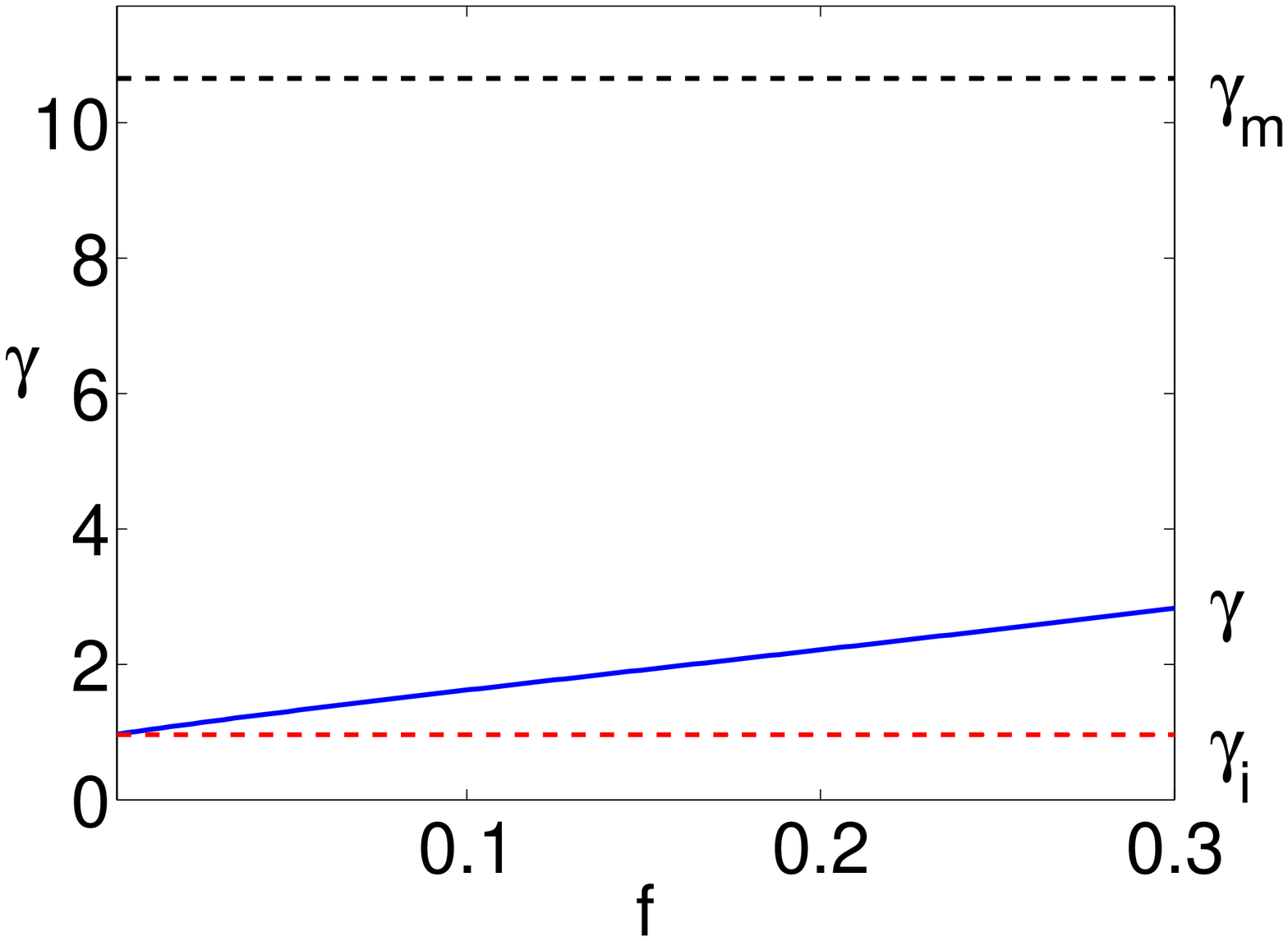}}
\caption{\label{fig:fig7all}Composite electrostriction $\mathrm{Re} (\gamma)$ from \eqref{eq:GammaMG} at $\lambda = 1550$ nm for an array of  \subref{fig:fig7} SiO$_2$ spheres embedded in a   Si matrix, and \subref{fig:fig7b} As$_2$S$_3$ spheres embedded  in a SiO$_2$ matrix. The electrostriction   for the inclusion $\gamma_\mathrm{i}$ (dashed black line) and  matrix  $\gamma_\mathrm{m}$ (dashed red line) materials are also given.}
\end{figure}

In Fig. \ref{fig:fig7} we show a cross section of the composite $\mathrm{Re} (\gamma)$ from Fig. \ref{fig:gsis} at $\lambda = 1550$ nm.  This gives confirmation  that complete suppression of electrostriction is achieved at    $f \approx 10\%$, and shows that we have sign-changing electrostriction from this metamaterial design. We note that the composite $\mathrm{Re} (\gamma)$ (blue curve) exceeds that of the constituent electrostriction values (dashed   curves) at  a filling fraction of $f = 16.6\%$, which is reminiscent of earlier work which showed the nonlinear parameters of composite materials can exceed the values of the constituents\cite{sipe1992nonlinear}, but we note that care must be taken as the dilute lattice assumption breaks down in this region of enhancement here.  An investigation using As$_2$S$_3$ spheres in a    Si matrix demonstrated an identical result to that   shown in Figures \ref{fig:gsis}, \ref{fig:esis} and  \ref{fig:fig7}, but at much lower filling fractions.  

To emphasise the result presented in Fig. \ref{fig:fig7}, we show the composite  $\gamma$ curve for  an array of chalcogenide (As$_2$S$_3$) spheres embedded in a silica matrix at $\lambda = 1550$ nm   in Fig. \ref{fig:fig7b}. This   shows a  simple    linear enhancement from   the background electrostriction, to a maximum realisable value of $ \gamma = 1.918$ at   the threshold of $f = 15\%$ (i.e. an enhancement factor of approximately 2).

 \section{Concluding remarks}\label{sec:concl} \noindent
 We have presented     an  analytical representation  for the electrostriction  of a composite material   by incorporating  the simplest and  analytically most  transparent  model  from effective index theory, the Maxwell--Garnett model,   to the problem of electrostriction.   
 
 We show that   expressions for the electrostriction of a composite material feature   artificial electrostriction terms, which  contribute to the enhancement or suppression of this material property, as observed for a selection of composites here. The presence of this term points towards   the possibility that large enhancements in $\gamma$, beyond both material values, could be achieved for more sophisticated metamaterial designs.     We also  show that    sign-switching electrostriction   is   achievable, and that resonant enhancements  in the electrostriction of metal-dielectric composites       are unrealistic, as they are associated with strong attenuation. Incorporating dispersive effects in the model is shown to give a small but non-negligible correction to estimates for the composite electrostriction.

 It is important to emphasise that this work is a first step in the study of the electrostriction of composites, and so other considerations such as thermally-induced electrostriction  and scattering losses, are not addressed here. These effects are more prominent  for high-intensity   wave problems, where more elaborate models are required to accurately evaluate the material response. Including the effect of shear stresses   will require a fully tensorial description of all stress fields, and is the next step in the development of the theory. Also, we   note that our estimate for the electrostriction of metals   is   a low-order  approximation, which requires experimental data for validation.  
   
As a final comment, we emphasise that other homogenisation procedures\cite{milton2002theory} can be used to   determine  the electrostriction for a periodic composite, which should remove several constraints of the present MG model, and    open  the way to investigations of  exciting metamaterial designs.  
  
 {\it Note added in proof.} Recently, we became aware of a preprint \cite{sun2015analytic} on a similar topic.

  \begin{acknowledgments}
This work was supported by the Australian Research Council (CUDOS Centre of Excellence, CE110001018).  
\end{acknowledgments}

\appendix

\section{Asymptotic analysis of the nondispersive composite electrostriction  }\label{app:1} \noindent
In this appendix we   examine several asymptotic limits for the composite electrostriction expression \eqref{eq:GammaMG}. First, assuming   $ \varepsilon_\mathrm{m}  \gg  \varepsilon_{\mathrm{i}} $, we obtain
\begin{multline}
\label{eq:gammaepslim1}
 \gamma \sim \frac{9 \beta_\rmi f}{\beta_\mathrm{c} (f+2)^2} \gamma_\rmi - \frac{2 \beta_\rmm (f-1)}{\beta_\mathrm{c} (f+2)} \gamma_\rmm + \frac{6 (\beta_\rmi - \beta_\rmm) f (1-f) \varepsilon_\rmm}{\beta_\mathrm{c} (f+2)^2}
  \end{multline}
where we have a persistent, but simplified, contribution from all terms in \eqref{eq:GammaMG}. By contrast, for $\varepsilon_\rmi \gg \varepsilon_\rmm$ we have	
\begin{equation}
\label{eq:gammaepslim2}
\gamma \sim  \frac{9 f \beta_\rmi}{(1-f)^2 \beta_\mathrm{c}}\left( \frac{ \varepsilon_\rmm^2}{ \varepsilon_\rmi^2} \right) \gamma_\rmi + \frac{\beta_\rmm (1+2f)}{(1-f) \beta_\mathrm{c}} \gamma_\rmm + \frac{3 \varepsilon_\rmm f (\beta_\rmm - \beta_\rmi)}{(1-f)\beta_\mathrm{c}}.
 \end{equation}
The limit  $\beta_\mathrm{m} \gg \beta_\mathrm{i}$ gives the asymptotic form
\begin{multline}
\gamma \sim  \left[ \frac{( \varepsilon_ {\mathrm{i}} + 2 \varepsilon_ {\mathrm{m}} )^2 + 2f ( \varepsilon_ {\mathrm{i}} - \varepsilon_ {\mathrm{m}} )^2 }{ \left[ ( \varepsilon_ {\mathrm{i}} + 2 \varepsilon_ {\mathrm{m}}) - ( \varepsilon_ {\mathrm{i}} - \varepsilon_ {\mathrm{m}}) f\right]^2}\right] \gamma_ {\mathrm{m}} \\ 
+ \frac{3  f  \varepsilon_ {\mathrm{m}} ( \varepsilon_ {\mathrm{i}} - \varepsilon_ {\mathrm{m}} ) ( \varepsilon_ {\mathrm{i}} + 2 \varepsilon_ {\mathrm{m}} ) }{ \left[ ( \varepsilon_ {\mathrm{i}} + 2 \varepsilon_ {\mathrm{m}}) - ( \varepsilon_ {\mathrm{i}} - \varepsilon_ {\mathrm{m}}) f\right]^2}, 
\end{multline}
and   $\beta_\mathrm{i} \gg \beta_\mathrm{m}$  leads to
\begin{multline}
\gamma \sim  \left[ \frac{ ( \varepsilon_ {\mathrm{i}} + 2 \varepsilon_ {\mathrm{m}} )  - ( \varepsilon_ {\mathrm{i}} - \varepsilon_ {\mathrm{m}}) }{ ( \varepsilon_ {\mathrm{i}} + 2 \varepsilon_ {\mathrm{m}}) -  ( \varepsilon_ {\mathrm{i}} - \varepsilon_ {\mathrm{m}}) f }\right]^2  \gamma_ {\mathrm{i}} \\ -  \frac{3 (1- f)  \varepsilon_ {\mathrm{m}} ( \varepsilon_ {\mathrm{i}} - \varepsilon_ {\mathrm{m}} ) ( \varepsilon_ {\mathrm{i}} + 2 \varepsilon_ {\mathrm{m}} ) }{ \left[ ( \varepsilon_ {\mathrm{i}} + 2 \varepsilon_ {\mathrm{m}}) - ( \varepsilon_ {\mathrm{i}} - \varepsilon_ {\mathrm{m}}) f\right]^2}. 
 \end{multline}
These difference in sign in the artificial electrostriction contributions above suggest that the relative magnitudes of $\beta$    are relevant in establishing whether enhanced or suppressed  electrostriction is observed.
 
\bibliography{electrostriction_bib}

\begin{thebibliography}{28}%
\makeatletter
\providecommand \@ifxundefined [1]{%
 \@ifx{#1\undefined}
}%
\providecommand \@ifnum [1]{%
 \ifnum #1\expandafter \@firstoftwo
 \else \expandafter \@secondoftwo
 \fi
}%
\providecommand \@ifx [1]{%
 \ifx #1\expandafter \@firstoftwo
 \else \expandafter \@secondoftwo
 \fi
}%
\providecommand \natexlab [1]{#1}%
\providecommand \enquote  [1]{``#1''}%
\providecommand \bibnamefont  [1]{#1}%
\providecommand \bibfnamefont [1]{#1}%
\providecommand \citenamefont [1]{#1}%
\providecommand \href@noop [0]{\@secondoftwo}%
\providecommand \href [0]{\begingroup \@sanitize@url \@href}%
\providecommand \@href[1]{\@@startlink{#1}\@@href}%
\providecommand \@@href[1]{\endgroup#1\@@endlink}%
\providecommand \@sanitize@url [0]{\catcode `\\12\catcode `\$12\catcode
  `\&12\catcode `\#12\catcode `\^12\catcode `\_12\catcode `\%12\relax}%
\providecommand \@@startlink[1]{}%
\providecommand \@@endlink[0]{}%
\providecommand \url  [0]{\begingroup\@sanitize@url \@url }%
\providecommand \@url [1]{\endgroup\@href {#1}{\urlprefix }}%
\providecommand \urlprefix  [0]{URL }%
\providecommand \Eprint [0]{\href }%
\providecommand \doibase [0]{http://dx.doi.org/}%
\providecommand \selectlanguage [0]{\@gobble}%
\providecommand \bibinfo  [0]{\@secondoftwo}%
\providecommand \bibfield  [0]{\@secondoftwo}%
\providecommand \translation [1]{[#1]}%
\providecommand \BibitemOpen [0]{}%
\providecommand \bibitemStop [0]{}%
\providecommand \bibitemNoStop [0]{.\EOS\space}%
\providecommand \EOS [0]{\spacefactor3000\relax}%
\providecommand \BibitemShut  [1]{\csname bibitem#1\endcsname}%
\let\auto@bib@innerbib\@empty
\bibitem [{\citenamefont {Eggleton}\ \emph {et~al.}(2013)\citenamefont
  {Eggleton}, \citenamefont {Poulton},\ and\ \citenamefont
  {Pant}}]{eggleton2013inducing}%
  \BibitemOpen
  \bibfield  {author} {\bibinfo {author} {\bibfnamefont {B.~J.}\ \bibnamefont
  {Eggleton}}, \bibinfo {author} {\bibfnamefont {C.~G.}\ \bibnamefont
  {Poulton}}, \ and\ \bibinfo {author} {\bibfnamefont {R.}~\bibnamefont
  {Pant}},\ }\href@noop {} {\bibfield  {journal} {\bibinfo  {journal} {Adv.
  Opt. Photonics}\ }\textbf {\bibinfo {volume} {5}},\ \bibinfo {pages} {536}
  (\bibinfo {year} {2013})}\BibitemShut {NoStop}%
\bibitem [{\citenamefont {Brillouin}(1922)}]{brillouin1922diffusion}%
  \BibitemOpen
  \bibfield  {author} {\bibinfo {author} {\bibfnamefont {L.}~\bibnamefont
  {Brillouin}},\ }\href@noop {} {\bibfield  {journal} {\bibinfo  {journal}
  {Ann. Phys.-Paris}\ }\textbf {\bibinfo {volume} {17}},\ \bibinfo {pages} {88}
  (\bibinfo {year} {1922})}\BibitemShut {NoStop}%
\bibitem [{\citenamefont {Mandelstam}(1926)}]{mandelstam1926light}%
  \BibitemOpen
  \bibfield  {author} {\bibinfo {author} {\bibfnamefont {L.}~\bibnamefont
  {Mandelstam}},\ }\href@noop {} {\bibfield  {journal} {\bibinfo  {journal}
  {Zh. Russ. Fiz-Khim. Ova}\ }\textbf {\bibinfo {volume} {58}},\ \bibinfo
  {pages} {381} (\bibinfo {year} {1926})}\BibitemShut {NoStop}%
\bibitem [{\citenamefont {Chiao}\ \emph {et~al.}(1964)\citenamefont {Chiao},
  \citenamefont {Garmire},\ and\ \citenamefont {Townes}}]{chiao1964self}%
  \BibitemOpen
  \bibfield  {author} {\bibinfo {author} {\bibfnamefont {R.~Y.}\ \bibnamefont
  {Chiao}}, \bibinfo {author} {\bibfnamefont {E.}~\bibnamefont {Garmire}}, \
  and\ \bibinfo {author} {\bibfnamefont {C.~H.}\ \bibnamefont {Townes}},\
  }\href@noop {} {\bibfield  {journal} {\bibinfo  {journal} {Phys. Rev. Lett.}\
  }\textbf {\bibinfo {volume} {13}},\ \bibinfo {pages} {479} (\bibinfo {year}
  {1964})}\BibitemShut {NoStop}%
\bibitem [{\citenamefont {Lapine}\ \emph {et~al.}(2014)\citenamefont {Lapine},
  \citenamefont {Shadrivov},\ and\ \citenamefont
  {Kivshar}}]{lapine2014colloquium}%
  \BibitemOpen
  \bibfield  {author} {\bibinfo {author} {\bibfnamefont {M.}~\bibnamefont
  {Lapine}}, \bibinfo {author} {\bibfnamefont {I.~V.}\ \bibnamefont
  {Shadrivov}}, \ and\ \bibinfo {author} {\bibfnamefont {Y.~S.}\ \bibnamefont
  {Kivshar}},\ }\href@noop {} {\bibfield  {journal} {\bibinfo  {journal} {Rev.
  Mod. Phys.}\ }\textbf {\bibinfo {volume} {86}},\ \bibinfo {pages} {1093}
  (\bibinfo {year} {2014})}\BibitemShut {NoStop}%
\bibitem [{\citenamefont {Agranovich}\ \emph {et~al.}(2004)\citenamefont
  {Agranovich}, \citenamefont {Shen}, \citenamefont {Baughman},\ and\
  \citenamefont {Zakhidov}}]{agranovich2004linear}%
  \BibitemOpen
  \bibfield  {author} {\bibinfo {author} {\bibfnamefont {V.~M.}\ \bibnamefont
  {Agranovich}}, \bibinfo {author} {\bibfnamefont {Y.~R.}\ \bibnamefont
  {Shen}}, \bibinfo {author} {\bibfnamefont {R.~H.}\ \bibnamefont {Baughman}},
  \ and\ \bibinfo {author} {\bibfnamefont {A.~A.}\ \bibnamefont {Zakhidov}},\
  }\href@noop {} {\bibfield  {journal} {\bibinfo  {journal} {Phys. Rev. B}\
  }\textbf {\bibinfo {volume} {69}},\ \bibinfo {pages} {165112} (\bibinfo
  {year} {2004})}\BibitemShut {NoStop}%
\bibitem [{\citenamefont {Segal}\ \emph {et~al.}(2015)\citenamefont {Segal},
  \citenamefont {Keren-Zur}, \citenamefont {Hendler},\ and\ \citenamefont
  {Ellenbogen}}]{segal2015controlling}%
  \BibitemOpen
  \bibfield  {author} {\bibinfo {author} {\bibfnamefont {N.}~\bibnamefont
  {Segal}}, \bibinfo {author} {\bibfnamefont {S.}~\bibnamefont {Keren-Zur}},
  \bibinfo {author} {\bibfnamefont {N.}~\bibnamefont {Hendler}}, \ and\
  \bibinfo {author} {\bibfnamefont {T.}~\bibnamefont {Ellenbogen}},\
  }\href@noop {} {\bibfield  {journal} {\bibinfo  {journal} {Nat. Photonics}\
  }\textbf {\bibinfo {volume} {9}},\ \bibinfo {pages} {180} (\bibinfo {year}
  {2015})}\BibitemShut {NoStop}%
\bibitem [{\citenamefont {Aspelmeyer}\ \emph {et~al.}(2014)\citenamefont
  {Aspelmeyer}, \citenamefont {Kippenberg},\ and\ \citenamefont
  {Marquardt}}]{aspelmeyer2014cavity}%
  \BibitemOpen
  \bibfield  {author} {\bibinfo {author} {\bibfnamefont {M.}~\bibnamefont
  {Aspelmeyer}}, \bibinfo {author} {\bibfnamefont {T.~J.}\ \bibnamefont
  {Kippenberg}}, \ and\ \bibinfo {author} {\bibfnamefont {F.}~\bibnamefont
  {Marquardt}},\ }\href@noop {} {\bibfield  {journal} {\bibinfo  {journal}
  {Rev. Mod. Phys.}\ }\textbf {\bibinfo {volume} {86}},\ \bibinfo {pages}
  {1391} (\bibinfo {year} {2014})}\BibitemShut {NoStop}%
\bibitem [{\citenamefont {Lapine}\ \emph {et~al.}(2012)\citenamefont {Lapine},
  \citenamefont {Shadrivov}, \citenamefont {Powell},\ and\ \citenamefont
  {Kivshar}}]{lapine2012magnetoelastic}%
  \BibitemOpen
  \bibfield  {author} {\bibinfo {author} {\bibfnamefont {M.}~\bibnamefont
  {Lapine}}, \bibinfo {author} {\bibfnamefont {I.~V.}\ \bibnamefont
  {Shadrivov}}, \bibinfo {author} {\bibfnamefont {D.~A.}\ \bibnamefont
  {Powell}}, \ and\ \bibinfo {author} {\bibfnamefont {Y.~S.}\ \bibnamefont
  {Kivshar}},\ }\href@noop {} {\bibfield  {journal} {\bibinfo  {journal} {Nat.
  Mater.}\ }\textbf {\bibinfo {volume} {11}},\ \bibinfo {pages} {30} (\bibinfo
  {year} {2012})}\BibitemShut {NoStop}%
\bibitem [{\citenamefont {Sipe}\ and\ \citenamefont
  {Boyd}(1992)}]{sipe1992nonlinear}%
  \BibitemOpen
  \bibfield  {author} {\bibinfo {author} {\bibfnamefont {J.~E.}\ \bibnamefont
  {Sipe}}\ and\ \bibinfo {author} {\bibfnamefont {R.~W.}\ \bibnamefont
  {Boyd}},\ }\href@noop {} {\bibfield  {journal} {\bibinfo  {journal} {Phys.
  Rev. A}\ }\textbf {\bibinfo {volume} {46}},\ \bibinfo {pages} {1614}
  (\bibinfo {year} {1992})}\BibitemShut {NoStop}%
\bibitem [{\citenamefont {Boyd}(2003)}]{boyd2003nonlinear}%
  \BibitemOpen
  \bibfield  {author} {\bibinfo {author} {\bibfnamefont {R.~W.}\ \bibnamefont
  {Boyd}},\ }\href@noop {} {\emph {\bibinfo {title} {Nonlinear optics}}},\
  \bibinfo {edition} {3rd}\ ed.\ (\bibinfo  {publisher} {Academic press},\
  \bibinfo {address} {London},\ \bibinfo {year} {2003})\BibitemShut {NoStop}%
\bibitem [{\citenamefont {Fabelinskii}(1968)}]{fabelinskii1968molecular}%
  \BibitemOpen
  \bibfield  {author} {\bibinfo {author} {\bibfnamefont {I.~L.}\ \bibnamefont
  {Fabelinskii}},\ }\href@noop {} {\emph {\bibinfo {title} {Molecular
  scattering of light}}}\ (\bibinfo  {publisher} {Springer},\ \bibinfo
  {address} {New York},\ \bibinfo {year} {1968})\BibitemShut {NoStop}%
\bibitem [{\citenamefont {Landau}\ \emph {et~al.}(1984)\citenamefont {Landau},
  \citenamefont {Lifshitz},\ and\ \citenamefont
  {Pitaevskii}}]{landau1984electrodynamics}%
  \BibitemOpen
  \bibfield  {author} {\bibinfo {author} {\bibfnamefont {L.~D.}\ \bibnamefont
  {Landau}}, \bibinfo {author} {\bibfnamefont {E.~M.}\ \bibnamefont
  {Lifshitz}}, \ and\ \bibinfo {author} {\bibfnamefont {L.~P.}\ \bibnamefont
  {Pitaevskii}},\ }\href@noop {} {\emph {\bibinfo {title} {Electrodynamics of
  continuous media}}},\ \bibinfo {edition} {2nd}\ ed.,\ Vol.~\bibinfo {volume}
  {8}\ (\bibinfo  {publisher} {Pergamon Press},\ \bibinfo {address} {Oxford},\
  \bibinfo {year} {1984})\BibitemShut {NoStop}%
\bibitem [{\citenamefont {Rakich}\ \emph {et~al.}(2010)\citenamefont {Rakich},
  \citenamefont {Davids},\ and\ \citenamefont {Wang}}]{rakich2010tailoring}%
  \BibitemOpen
  \bibfield  {author} {\bibinfo {author} {\bibfnamefont {P.~T.}\ \bibnamefont
  {Rakich}}, \bibinfo {author} {\bibfnamefont {P.}~\bibnamefont {Davids}}, \
  and\ \bibinfo {author} {\bibfnamefont {Z.}~\bibnamefont {Wang}},\ }\href@noop
  {} {\bibfield  {journal} {\bibinfo  {journal} {Opt. Expr.}\ }\textbf
  {\bibinfo {volume} {18}},\ \bibinfo {pages} {14439} (\bibinfo {year}
  {2010})}\BibitemShut {NoStop}%
\bibitem [{\citenamefont {Stratton}(2007)}]{stratton2007electromagnetic}%
  \BibitemOpen
  \bibfield  {author} {\bibinfo {author} {\bibfnamefont {J.~A.}\ \bibnamefont
  {Stratton}},\ }\href@noop {} {\emph {\bibinfo {title} {Electromagnetic
  theory}}}\ (\bibinfo  {publisher} {John Wiley \& Sons},\ \bibinfo {address}
  {Hoboken},\ \bibinfo {year} {2007})\BibitemShut {NoStop}%
\bibitem [{\citenamefont {Nelson}\ and\ \citenamefont
  {Lax}(1971)}]{nelson1971theory}%
  \BibitemOpen
  \bibfield  {author} {\bibinfo {author} {\bibfnamefont {D.~F.}\ \bibnamefont
  {Nelson}}\ and\ \bibinfo {author} {\bibfnamefont {M.}~\bibnamefont {Lax}},\
  }\href@noop {} {\bibfield  {journal} {\bibinfo  {journal} {Phys. Rev. B}\
  }\textbf {\bibinfo {volume} {3}},\ \bibinfo {pages} {2778} (\bibinfo {year}
  {1971})}\BibitemShut {NoStop}%
\bibitem [{\citenamefont {Mueller}(1935)}]{mueller1935theory}%
  \BibitemOpen
  \bibfield  {author} {\bibinfo {author} {\bibfnamefont {H.}~\bibnamefont
  {Mueller}},\ }\href@noop {} {\bibfield  {journal} {\bibinfo  {journal} {Phys.
  Rev.}\ }\textbf {\bibinfo {volume} {47}},\ \bibinfo {pages} {947} (\bibinfo
  {year} {1935})}\BibitemShut {NoStop}%
\bibitem [{\citenamefont {Melloni}\ \emph {et~al.}(1998)\citenamefont
  {Melloni}, \citenamefont {Frasca}, \citenamefont {Garavaglia}, \citenamefont
  {Tonini},\ and\ \citenamefont {Martinelli}}]{melloni1998direct}%
  \BibitemOpen
  \bibfield  {author} {\bibinfo {author} {\bibfnamefont {A.}~\bibnamefont
  {Melloni}}, \bibinfo {author} {\bibfnamefont {M.}~\bibnamefont {Frasca}},
  \bibinfo {author} {\bibfnamefont {A.}~\bibnamefont {Garavaglia}}, \bibinfo
  {author} {\bibfnamefont {A.}~\bibnamefont {Tonini}}, \ and\ \bibinfo {author}
  {\bibfnamefont {M.}~\bibnamefont {Martinelli}},\ }\href@noop {} {\bibfield
  {journal} {\bibinfo  {journal} {Opt. Lett.}\ }\textbf {\bibinfo {volume}
  {23}},\ \bibinfo {pages} {691} (\bibinfo {year} {1998})}\BibitemShut
  {NoStop}%
\bibitem [{\citenamefont {Stakhin}(1998)}]{stakhin1998electrostriction}%
  \BibitemOpen
  \bibfield  {author} {\bibinfo {author} {\bibfnamefont {N.~A.}\ \bibnamefont
  {Stakhin}},\ }\href@noop {} {\bibfield  {journal} {\bibinfo  {journal} {Russ.
  Phys. J.}\ }\textbf {\bibinfo {volume} {41}},\ \bibinfo {pages} {1107}
  (\bibinfo {year} {1998})}\BibitemShut {NoStop}%
\bibitem [{\citenamefont {Weber}(2002)}]{weber2002handbook}%
  \BibitemOpen
  \bibfield  {author} {\bibinfo {author} {\bibfnamefont {M.~J.}\ \bibnamefont
  {Weber}},\ }\href@noop {} {\emph {\bibinfo {title} {Handbook of optical
  materials}}}\ (\bibinfo  {publisher} {CRC press},\ \bibinfo {address} {Boca
  Raton},\ \bibinfo {year} {2002})\BibitemShut {NoStop}%
\bibitem [{\citenamefont {Biegelsen}(1974)}]{biegelsen1974photoelastic}%
  \BibitemOpen
  \bibfield  {author} {\bibinfo {author} {\bibfnamefont {D.~K.}\ \bibnamefont
  {Biegelsen}},\ }\href@noop {} {\bibfield  {journal} {\bibinfo  {journal}
  {Phys. Rev. Lett.}\ }\textbf {\bibinfo {volume} {32}},\ \bibinfo {pages}
  {1196} (\bibinfo {year} {1974})}\BibitemShut {NoStop}%
\bibitem [{\citenamefont {Bohren}\ and\ \citenamefont
  {Huffman}(2008)}]{bohren2008absorption}%
  \BibitemOpen
  \bibfield  {author} {\bibinfo {author} {\bibfnamefont {C.~F.}\ \bibnamefont
  {Bohren}}\ and\ \bibinfo {author} {\bibfnamefont {D.~R.}\ \bibnamefont
  {Huffman}},\ }\href@noop {} {\emph {\bibinfo {title} {Absorption and
  scattering of light by small particles}}}\ (\bibinfo  {publisher} {John Wiley
  \& Sons},\ \bibinfo {address} {New York},\ \bibinfo {year}
  {2008})\BibitemShut {NoStop}%
\bibitem [{\citenamefont {Milton}(2002)}]{milton2002theory}%
  \BibitemOpen
  \bibfield  {author} {\bibinfo {author} {\bibfnamefont {G.~W.}\ \bibnamefont
  {Milton}},\ }\href@noop {} {\emph {\bibinfo {title} {The theory of
  composites}}}\ (\bibinfo  {publisher} {Cambridge University Press},\ \bibinfo
  {address} {New York},\ \bibinfo {year} {2002})\BibitemShut {NoStop}%
\bibitem [{\citenamefont {Ordal}\ \emph {et~al.}(1985)\citenamefont {Ordal},
  \citenamefont {Bell}, \citenamefont {Alexander~Jr}, \citenamefont {Long},\
  and\ \citenamefont {Querry}}]{ordal1985optical}%
  \BibitemOpen
  \bibfield  {author} {\bibinfo {author} {\bibfnamefont {M.~A.}\ \bibnamefont
  {Ordal}}, \bibinfo {author} {\bibfnamefont {R.~J.}\ \bibnamefont {Bell}},
  \bibinfo {author} {\bibfnamefont {R.~W.}\ \bibnamefont {Alexander~Jr}},
  \bibinfo {author} {\bibfnamefont {L.~L.}\ \bibnamefont {Long}}, \ and\
  \bibinfo {author} {\bibfnamefont {M.~R.}\ \bibnamefont {Querry}},\
  }\href@noop {} {\bibfield  {journal} {\bibinfo  {journal} {App. Opt.}\
  }\textbf {\bibinfo {volume} {24}},\ \bibinfo {pages} {4493} (\bibinfo {year}
  {1985})}\BibitemShut {NoStop}%
\bibitem [{\citenamefont {Malitson}(1965)}]{malitson1965interspecimen}%
  \BibitemOpen
  \bibfield  {author} {\bibinfo {author} {\bibfnamefont {I.~H.}\ \bibnamefont
  {Malitson}},\ }\href@noop {} {\bibfield  {journal} {\bibinfo  {journal} {J.
  Opt. Soc. Am.}\ }\textbf {\bibinfo {volume} {55}},\ \bibinfo {pages} {1205}
  (\bibinfo {year} {1965})}\BibitemShut {NoStop}%
\bibitem [{amo(2012)}]{amorphmat}%
  \BibitemOpen
  \href@noop {} {\enquote {\bibinfo {title} {Amtir-6 information},}\ }\bibinfo
  {howpublished}
  {\url{http://www.amorphousmaterials.com/app/download/6552919404/AMTIR-6+Information.pdf}}
  (\bibinfo {year} {2012}),\ \bibinfo {note} {accessed: 2015-02-12}\BibitemShut
  {NoStop}%
\bibitem [{\citenamefont {Green}(2008)}]{green2008self}%
  \BibitemOpen
  \bibfield  {author} {\bibinfo {author} {\bibfnamefont {M.~A.}\ \bibnamefont
  {Green}},\ }\href@noop {} {\bibfield  {journal} {\bibinfo  {journal} {Sol.
  Energ. Mat. Sol. C}\ }\textbf {\bibinfo {volume} {92}},\ \bibinfo {pages}
  {1305} (\bibinfo {year} {2008})}\BibitemShut {NoStop}%
\bibitem [{\citenamefont {Sun}\ \emph {et~al.}(2015)\citenamefont {Sun},
  \citenamefont {Ng}, \citenamefont {Zhou},\ and\ \citenamefont
  {Chan}}]{sun2015analytic}%
  \BibitemOpen
  \bibfield  {author} {\bibinfo {author} {\bibfnamefont {W.}~\bibnamefont
  {Sun}}, \bibinfo {author} {\bibfnamefont {J.}~\bibnamefont {Ng}}, \bibinfo
  {author} {\bibfnamefont {L.}~\bibnamefont {Zhou}}, \ and\ \bibinfo {author}
  {\bibfnamefont {C.}~\bibnamefont {Chan}},\ }\href@noop {} {\bibfield
  {journal} {\bibinfo  {journal} {arXiv preprint arXiv:1504.06437}\ } (\bibinfo
  {year} {2015})}\BibitemShut {NoStop}%
\end{thebibliography}%

\end{document}